# Layer-dependent Ferromagnetism in a van der Waals Crystal down to the Monolayer Limit

Bevin Huang[1†], Genevieve Clark[2†], Efren Navarro-Moratalla[3†], Dahlia R. Klein[3], Ran Cheng[4], Kyle L. Seyler[1], Ding Zhong[1], Emma Schmidgall[1], Michael A. McGuire[5], David H. Cobden[1], Wang Yao[6], Di Xiao[4], Pablo Jarillo-Herrero[3*], Xiaodong Xu[1,2*]

[1]Department of Physics, University of Washington, Seattle, Washington 98195, USA
[2]Department of Materials Science and Engineering, University of Washington, Seattle, Washington 98195, USA
[3]Department of Physics, Massachusetts Institute of Technology, Cambridge, Massachusetts 02139, USA
[4]Department of Physics, Carnegie Mellon University, Pittsburgh, Pennsylvania 15213, USA
[5]Materials Science and Technology Division, Oak Ridge National Laboratory, Oak Ridge, Tennessee, 37831, USA
[6]Department of Physics and Center of Theoretical and Computational Physics, University of Hong Kong, Hong Kong, China

[†]These authors contributed equally to this work.
[*]Correspondence to: xuxd@uw.edu, pjarillo@mit.edu

**Abstract:** Since the celebrated discovery of graphene[1,2], the family of two-dimensional (2D) materials has grown to encompass a broad range of electronic properties. Recent additions include spin-valley coupled semiconductors[3], Ising superconductors[4-6] that can be tuned into a quantum metal[7], possible Mott insulators with tunable charge-density waves[8], and topological semi-metals with edge transport[9,10]. Despite this progress, there is still no 2D crystal with intrinsic magnetism[11-16], which would be useful for many technologies such as sensing, information, and data storage[17]. Theoretically, magnetic order is prohibited in the 2D isotropic Heisenberg model at finite temperatures by the Mermin-Wagner theorem[18]. However, magnetic anisotropy removes this restriction and enables, for instance, the occurrence of 2D Ising ferromagnetism. Here, we use magneto-optical Kerr effect (MOKE) microscopy to demonstrate that *monolayer* chromium triiodide ($CrI_3$) is an Ising ferromagnet with out-of-plane spin orientation. Its Curie temperature of 45 K is only slightly lower than the 61 K of the bulk crystal, consistent with a weak interlayer coupling. Moreover, our studies suggest a layer-dependent magnetic phase transition, showcasing the hallmark thickness-dependent physical properties typical of van der Waals crystals[19-21]. Remarkably, bilayer $CrI_3$ displays suppressed magnetization with a metamagnetic effect[22], while in trilayer the interlayer ferromagnetism observed in the bulk crystal is restored. Our work creates opportunities for studying magnetism by harnessing the unique features of atomically-thin materials, such as electrical control for realizing magnetoelectronics[13,23], and van der Waals engineering for novel interface phenomena[17].

**Main Text:**

Magnetic anisotropy is an important requirement for realizing 2D magnetism. In ultrathin metallic films, an easy-axis can originate from symmetry reduction at the interface/surface, which hinges on substrate properties and interface quality[24-26]. In contrast, most van der Waals magnets have an intrinsic magnetocrystalline anisotropy due to the reduced crystal symmetry of their layered structures. This offers the coveted possibility to retain a magnetic ground state in the monolayer limit. In addition to studying magnetism in naturally formed crystals in the true 2D limit, layered magnets provide a platform for studying the thickness dependence of magnetism in isolated single crystals where the interaction with the underlying substrate is weak. Namely, the covalently bonded van der Waals layers prevent complex magnetization reorientations induced by epitaxial lattice reconstruction and strain[23]. For layered materials, these advantages come at a low fabrication cost, since the micromechanical exfoliation technique[27] is much simpler than conventional approaches requiring sputtering or sophisticated molecular beam epitaxy.

A variety of layered magnetic compounds have recently drawn increased interest due to the possibility of retaining their magnetic properties down to monolayer thickness[12-15,21]. Recent Raman studies suggest ferromagnetic ordering in few-layer $Cr_2Ge_2Te_6$ and antiferromagnetic ordering in monolayer $FePS_3$[28]. However, no evidence yet exists for ferromagnetism persisting down to the *monolayer* limit. One promising candidate is bulk crystalline $CrI_3$. It shows layered Ising ferromagnetism below a Curie temperature ($T_c$) of 61 K with an out-of-plane easy-axis (Figs. 1a&b)[29,30]. Given its van der Waals nature, we expect magnetocrystalline anisotropy, which can lift the Mermin-Wagner restriction, to stabilize long-range ferromagnetic ordering even in a monolayer.

In our experiment, we obtained atomically-thin $CrI_3$ flakes by mechanical exfoliation of bulk crystals onto oxidized silicon substrates (See methods for $CrI_3$ crystal growth[11] and fabrication details). Given the reactivity of $CrI_3$ flakes, sample preparation was carried out in a glove box under an inert atmosphere. We mainly employed optical contrast based on the pixel RGB value to index the number of layers in a flake[31]. Figure 1c is an optical micrograph of a typical multi-step $CrI_3$ flake on a 285 nm $SiO_2$/Si substrate, showing regions ranging from 1 to 6 layers. Figure 1d shows an optical contrast map of the same region illuminated by 631 nm filtered light. The extracted optical contrast as a function of layer thickness is in good agreement with models based on the Fresnel equations (Fig. 1e)[31]. To accurately determine the correspondence between optical contrast and flake thickness, we also measured the thickness of $CrI_3$ flakes by atomic force microscopy, determined to be 0.7 nm/layer, after encapsulation with few-layer graphene (See Extended Data Fig. 1).

To probe the magnetic order, we employed polar MOKE measurements as a function of applied external magnetic field perpendicular to the sample plane (Faraday geometry). This design is sensitive to out-of-plane magnetization, and can detect small Kerr rotations, $\theta_K$, of linearly-polarized light down to 100 µrad using an AC polarization modulation technique[32] as laid out in Extended Data Fig. 2. All optical measurements were carried out using a 633 nm HeNe laser and at a temperature of 15 K, unless otherwise specified. Figure 1f illustrates the MOKE signal from a thin bulk flake of $CrI_3$. The observed hysteresis curve and remanent $\theta_K$ at zero magnetic field $\mu_oH = 0$ T are hallmarks of ferromagnetic ordering, consistent with its bulk ferromagnetism with out-of-plane magnetization. Here, $\mu_o$ is the magnetic constant. The negative remanent $\theta_K$ when approaching zero field from a positive external field is a consequence of thin-film interference from reflections at the $CrI_3$-$SiO_2$ and $SiO_2$-Si interfaces (Supplementary materials).

Remarkably, the ferromagnetic ordering remains in the monolayer limit. Figure 2a shows $\theta_K$ as a function of $\mu_oH$ for a monolayer $CrI_3$ flake (Fig. 2a inset). A single hysteresis loop in $\theta_K$ centered around $\mu_oH = 0$ T, with a nonzero remanent Kerr rotation, demonstrates out-of-plane spin polarization. This implies Ising ferromagnetism in monolayer $CrI_3$. As expected, $\theta_K$ is independent of the excitation power (Fig. 2b). In the rest of the letter, all data are taken with an excitation power of 10 µW. We have measured a total of 12 monolayer samples, which show similar MOKE behavior with consistent remanent $\theta_K$ values of about $5 \pm 2$ mrad at $\mu_oH = 0$ T (Extended Data Fig. 3a). The coercive field ($\mu_oH_c$), which is ~50 mT for the sample in Fig. 2a, can vary between samples due to the formation of domain structures in some samples.

Figure 2c shows spatial maps of $\theta_K$ for another monolayer, taken at selected magnetic field values. After cooling the sample from above $T_c$ at $\mu_oH = 0$ T, the entire monolayer is spontaneously magnetized (defined as spin down, blue). As the field is increased to 0.15 T, the magnetization in the upper half of the flake switches direction (now spin up, red). As the field is further increased to 0.3 T, the lower half of the monolayer flips and the entire flake becomes spin up, parallel to $\mu_oH$. This observation of micron-scale lateral domains suggests different values of coercivity in each domain. Indeed, magnetic field sweeps ($\theta_K$ vs $\mu_oH$) taken at discrete points ranging across both domains (Fig. 2d) show the difference in coercive field between the upper and lower half of the monolayer. Sweeps taken only on the upper domain, marked by a blue circle, show a much narrower hysteresis loop (about 50 mT) than sweeps from spots on the lower domain (orange and purple circles, about 200 mT). When the beam spot is centered between the two domains, contributions from both can be seen in the resulting hysteresis loop (green circle), a consequence of the ~1 µm beam spot illuminating both domains.

To determine the monolayer $T_c$, we perform an analysis of the irreversible field cooled (FC) and zero-field cooled (ZFC) Kerr signal. ZFC sweeps were performed by cooling the sample in zero field and measuring $\theta_K$ while warming in a small magnetic field ($\mu_oH = 0.15$ T). Following domain disappearance at a temperature well above $T_c$ (90 K), the sample is cooled back down in the presence of the same external magnetic field. Thermomagnetic irreversibility can be observed below $T_c$, at which point the ZFC sweep and the FC sweep diverge as illustrated in Fig. 2e. We measured the average $T_c$ for the monolayer samples to be 45 K, slightly lower than the value (61 K) for bulk samples.

The layered structure of $CrI_3$ provides a unique opportunity to investigate ferromagnetism as a function of layer thickness. Figures 3a-c show $\theta_K$ vs $\mu_oH$ for representative 1-3 layer $CrI_3$ samples. All measured monolayer and trilayer samples consistently show ferromagnetic behavior with a single hysteresis loop centered at $\mu_oH = 0$ T (Figs. 3a&c, and Extended Data Fig. 3). Both remanent and saturation values of $\theta_K$ for trilayers are about $50 \pm 10$ mrad, which is an order of magnitude larger than for monolayers. This drastic change of $\theta_K$ from monolayer to trilayer may be due to a layer-dependent electronic structure, leading to weaker optical resonance effects at 633 nm for monolayer than for trilayer (Extended Data Fig. 4). We find that for trilayers and thin bulk samples, $T_c$ is consistent with the bulk value of 61 K. The relatively small decrease of $T_c$ from bulk to few-layer and monolayer samples suggests that interlayer interactions do not dominate the ferromagnetic ordering in $CrI_3$. Compared with metallic magnetic thin films whose magnetic properties strongly depend on the underlying substrate[33,34], the weak layer-dependent $T_c$ also implies a negligible substrate effect on the ferromagnetic phenomena in atomically-thin $CrI_3$. As such, exfoliated $CrI_3$ of all thicknesses can be regarded as isolated single crystals.

A further observation is that bilayer $CrI_3$ shows a markedly different magnetic behavior (Fig. 3b). For all ten bilayer samples measured, the MOKE signal is strongly suppressed, with $\theta_K$ approaching zero (subject to slight variation between samples, Extended Data Fig. 3b) at field values between ±0.65 T. This observation implies a compensation of the out-of-plane magnetization. Upon crossing a critical field, $\theta_K$ shows a sharp jump, depicting a sudden recovery of the out-of-plane co-parallel orientation of the spins. This new magnetic state has an order of magnitude larger saturation $\theta_K$ (40 ± 10 mrad) than monolayer samples, and slightly smaller than for trilayers.

The suppression of Kerr signal at zero magnetic field demonstrates that the ground state has zero out-of-plane magnetization. The plateau behavior of the M-H curve further implies that there are no in-plane spin components; otherwise, one would expect a gradual increase of the MOKE signal with increasing perpendicular magnetic field. Rather, our observation suggests that each individual layer is ferromagnetically ordered (out-of-plane) while the interlayer coupling is antiferromagnetic. In this case, the strength of the interlayer coupling determines the field at which jumps between different plateaus occur, around ±0.65 T. Although the detailed mechanism of this coupling remains an open question, it may originate from the interplay between superexchange and magnetic dipole-dipole interactions. Another bilayer feature distinct from that of the monolayers is the vanishingly small hysteresis around the jumps, suggesting negligible net perpendicular anisotropy. A possible interpretation is that the shape anisotropy (which prefers in-plane spin orientation) nearly compensates the intrinsic magnetocrystalline anisotropy (which prefers out-of-plane spin orientation) so that the overall anisotropy is close to zero.

The insets in Fig. 3b display the layer-by-layer switching behavior that leads to plausible magnetic ground states of bilayer $CrI_3$. When the magnetic field is within ±0.65 T, the magnetization of the two layers are oppositely oriented to one another. Thus, the net magnetization vanishes and bilayer $CrI_3$ behaves as an antiferromagnet with an exchange field of about 0.65 T. When $|\mu_o H| > 0.65$ T, magnetization in one layer flips to align with the external magnetic field and restores out-of-plane magnetization, giving rise to the large MOKE signal. At around $|\mu_o H| = 0.65$ T, the MOKE signal sharply increases from near zero to its saturation value within about 100 mT, suggesting an abrupt increase of out-of-plane magnetization by a small change of magnetic field. Such behavior is indicative of metamagnetism, the magnetic field-driven transition from antiferromagnetic ordering to a fully spin-polarized state[22].

In summary, we demonstrated 2D ferromagnetism in exfoliated monolayer $CrI_3$. The observed monolayer $T_c$ suggests that these 2D magnets are weakly coupled to their substrates and can be regarded as isolated magnets. This is in distinct contrast with conventional metallic monolayer films whose magnetism is strongly affected by substrate coupling. We also observed strong evidence for layer-dependent magnetic phase transitions, from ferromagnetism in the monolayer, to antiferromagnetism in bilayer, and back to ferromagnetism in the trilayer and bulk. We envision that the demonstration of intrinsic ferromagnetism in monolayer $CrI_3$ as well as its layer dependent magnetic behavior provide fascinating opportunities for the investigation of novel quantum phenomena, such as topological effects in novel hybrid superconducting-ferromagnetic van der Waals heterostructures, as well as the engineering of new magneto-optoelectronic devices, such as ferromagnetic light emitters.

**Methods:**

*Growth of chromium(III) iodide bulk crystals*: chromium powder (99.5%, Sigma-Aldrich) and anhydrous iodine beads (99.999%, same supplier) were mixed in a 1 : 3 ratio inside a glove box with an argon atmosphere. 1.5 g of the mixture was then loaded in a silica ampoule (16 mm of inner diameter, 19 mm of outer diameter and 550 mm in length). The ampoule was extracted from the glove box and immediately evacuated to a pressure of approximately $10^{-4}$ Torr. Once at that pressure, the closed end was dipped in liquid nitrogen to prevent the sublimation of the iodine beads. The ampoule was then flame-sealed under dynamic vacuum and finally placed inside the three-zone furnace. Following an inverted gradient step of several hours, the crystals were grown over a period of 7 days with source zone at 650 ºC (containing the solid mixture), middle growth zone at 550 ºC and third zone at 600 ºC. Crystals formed both in the source (lustrous hexagonal platelets of several millimeters in size) and the middle (millimeter-long ribbon-like flakes) zones. The crystals were extracted from the ampoule in an argon atmosphere and stored in anhydrous conditions. The I : Cr elemental ratio was verified to be 2.8 ± 0.2 in several crystals by energy-dispersive X-Ray microanalysis performed on individual crystals in a Zeiss Merlin High-resolution SEM equipped with an EDS probe. In order to confirm the crystallographic phase of the material, a few single crystals were ground, loaded into a 0.3 mm outer diameter capillary and mounted on a Rigaku Smartlab Multipurpose Diffractometer setup in converging beam configuration with a D/teX detector. The room temperature X-Ray diffraction patterns of both the ribbon-like and hexagonal platelets were identical and consistent with the high-temperature monoclinic $AlCl_3$-type structure ($C2/m$) reported for $CrI_3$, with indexed unit cell parameters of (Å): $a$ = 6.8735(2), $b$ = 11.8859(3), $c$ = 6.9944(1) and $\beta$ = 108.535(2) (Le-Bail refinement, $R_{Bragg}$ = 5.27%). SQUID magnetometry performed on the single crystals depicts a $T_c$ of 61 K and a saturation magnetization of 3 $\mu_B$ per Cr, also in agreement with the values reported in the literature (see Supplementary Materials).

*Encapsulation of samples for AFM and measurement:* $CrI_3$ samples exfoliated under an inert atmosphere were encapsulated to preserve the $CrI_3$ flakes during AFM studies in ambient conditions. Using an all-dry viscoelastic stamping technique inside the glovebox, $CrI_3$ flakes were sandwiched between two layers of ~ 5 nm graphite to prevent the reaction with oxygen and moisture. Encapsulated $CrI_3$ flakes could then be safely removed from the glove box for further study. AFM of graphite encapsulated $CrI_3$ flakes was measured using a Bruker Dimension Edge atomic force microscope in tapping mode. Extended data Figure 1 shows AFM results for bilayer and tri-layer $CrI_3$ flakes encapsulated in ~ 5 nm graphite layers. Their corresponding MOKE data are consistent with bare samples.

*MOKE:* Power-stabilized light from a 633 nm HeNe laser source was linearly polarized at 45 degrees to the photoelastic modulator (PEM) slow axis (Extended Data Figure 2). Transmitting through the PEM, the light was sinusoidally phase-modulated at 50.1 kHz, with a maximum retardance of $\lambda/4$. Upon phase modulation, the light was focused down onto the sample at normal incidence using an aspheric lens. At any point in time when the light was not circularly polarized, reflection off a magnetic sample would rotate the polarization axis (the major axis for elliptically polarized light) by what we define as the Kerr rotation, $\theta_K$. This reflection was then separated from the incidence path via a laser line non-polarizing beamsplitter cube and projected onto the PEM slow axis using a second polarizer. For $\theta_K = 0$, the slow axis component remains constant for all polarizations and hence is not time-dependent. With a non-zero $\theta_K$, however, the slow axis component depends on the polarization and oscillates at twice the modulation frequency, 100.2 kHz, with an amplitude that is proportional to the Kerr rotation. Therefore, to obtain $\theta_K$, the reflection was detected using an amplified photodiode and two lock-in amplifiers: one tuned at 100.2 kHz to detect the Kerr rotation, and one tuned at the chopper frequency, 800 Hz, to normalize the Kerr signal to laser intensity fluctuations and the reflectivity of the sample.

**Acknowledgements:** Work at the University of Washington was mainly supported by the Department of Energy, Basic Energy Sciences, Materials Sciences and Engineering Division (DE-SC0008145 and SC0012509), and University of Washington Innovation Award. Work at MIT has been supported by the supported by the Center for Integrated Quantum Materials under NSF grant DMR-1231319 as well as the Gordon and Betty Moore Foundation's EPiQS Initiative through Grant GBMF4541 to PJH. Device fabrication has been partly supported by the Center for Excitonics, an Energy Frontier Research Center funded by the US Department of Energy (DOE), Office of Science, Office of Basic Energy Sciences under Award Number DESC0001088. DC's contribution is supported by DE-SC0002197. WY is supported by the Croucher Foundation (Croucher Innovation Award), the RGC of Hong Kong (HKU17305914P), and the HKU ORA. Work at ORNL (MAM) was supported by the US Department of Energy, Office of Science, Basic Energy Sciences, Materials Sciences and Engineering Division. XX and DX acknowledge the support a Cottrell Scholar Award. XX acknowledges the support from the State of Washington funded Clean Energy Institute and from the Boeing Distinguished Professorship in Physics.

**Author Contributions:** XX and PJH supervised the project. ENM and MAM synthesized and characterized the bulk $CrI_3$ crystal. ENM and DRK fabricated the samples and analyzed the layer thickness, assisted by GC and BH. BH built the MOKE setup with the help from ES and DZ. GC and BH performed the MOKE measurements, assisted by KS and ENM. RC, DX and WY provided the theoretical support. BH, GC, ENM, XX, PJH, DX and DC wrote the paper with input from all authors. All authors discussed the results.

**Figures:**

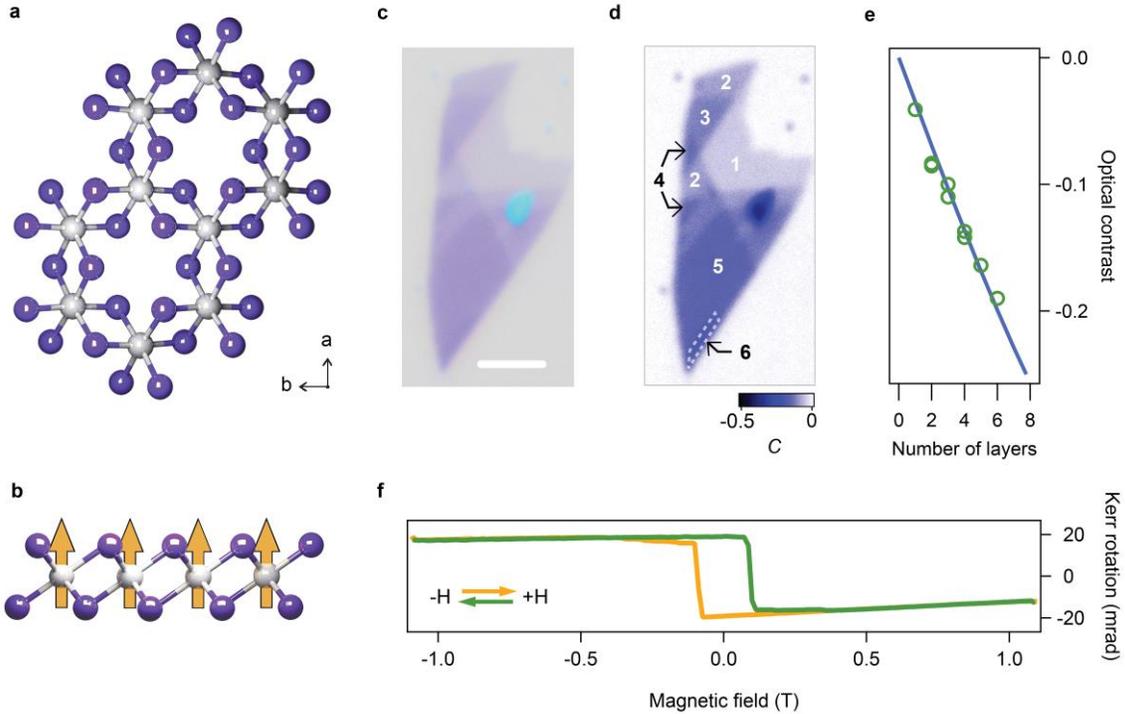

**Figure 1 | Crystal Structure, layer thickness identification, and MOKE of bulk $CrI_3$. a**, View of the in-plane atomic lattice of a single $CrI_3$ layer. Gray and purple balls represent Cr and I atoms, respectively. The $Cr^{3+}$ ions are coordinated to six $I^-$ ions to form edge-sharing octahedra arranged in a hexagonal honeycomb lattice. **b**, Out-of-plane view of the same $CrI_3$ structure depicting the Ising spin orientation. Atomic positions for these cartoons have been extracted from the crystal structure of bulk $CrI_3$ (crystallographic axis kept in the figure for reference). **c**, Optical microscope image of a representative $CrI_3$ flake and **d**, the calculated optical contrast map of the same flake with a 631 nm optical filter. The scale bar is 3 μm. **e**, Averaged optical contrast of the steps of sample with different number of layers (circles) fitted by a model based in Fresnel's equations (solid line). **f**, Polar MOKE signal of a thin bulk $CrI_3$ crystal.

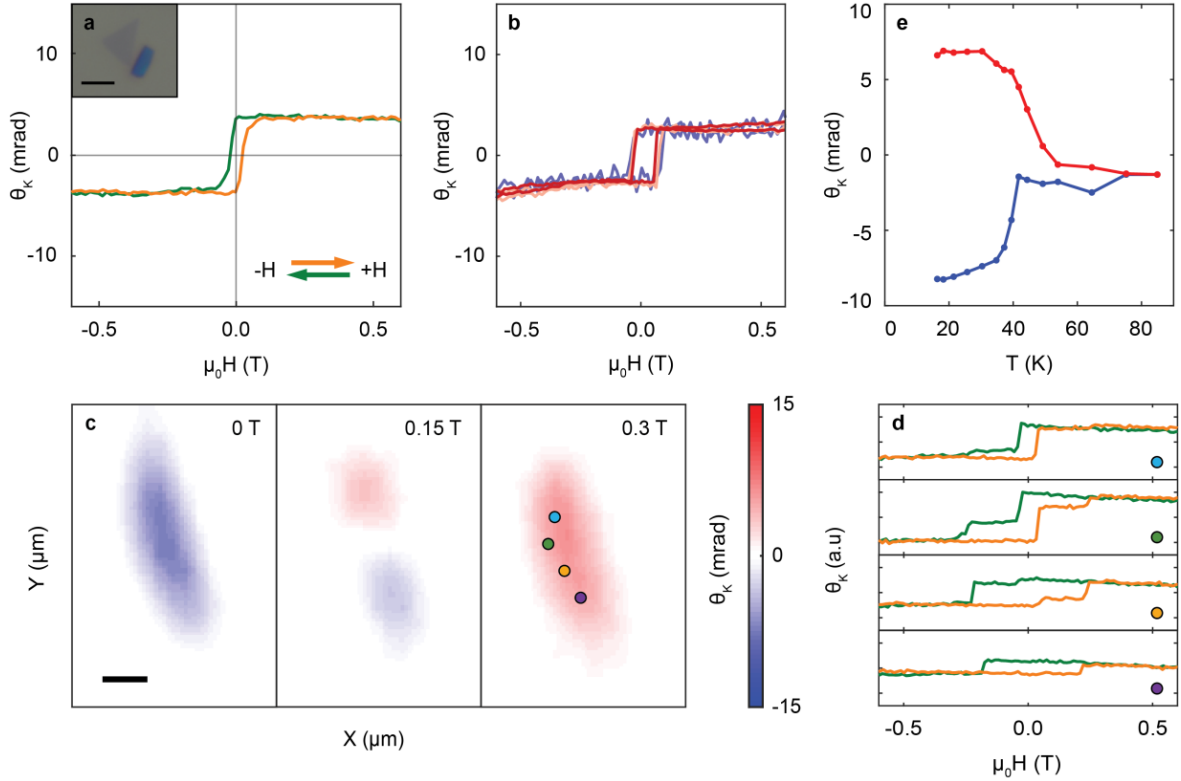

**Figure 2 | MOKE measurements of monolayer CrI$_3$. a**, Polar MOKE signal for a CrI$_3$ monolayer. Inset shows the optical image of an isolated monolayer. The scale bar is 2 μm. **b**, Power dependence of MOKE signal taken at incident powers of 3 μW (blue), 10 μW (pink), and 30 μW (red). **c,** MOKE maps at μ$_0$H = 0 T, 0.15 T, and 0.3 T on a different monolayer. The scale bar is 1 μm. **d,** θ$_K$ vs. μ$_0$H sweeps taken at four points marked by dots on the μ$_0$H = 0.3 T map in **c. e**, Temperature dependence of MOKE signal with the sample initially cooled at μ$_0$H = 0 T (blue) and 0.15 T (red).

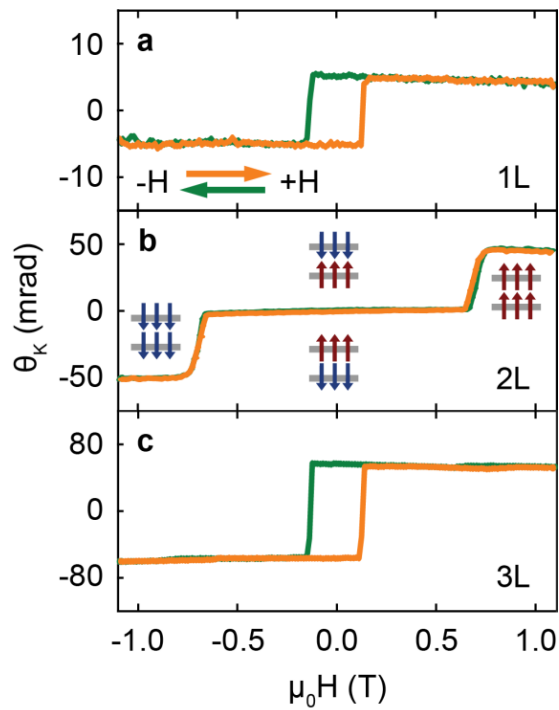

**Figure 3 | Layer-dependent magnetic ordering in atomically-thin CrI$_3$.** (**a**), MOKE signal on a monolayer CrI$_3$ flake, showing hysteresis in the Kerr rotation as a function of applied magnetic field, indicative of ferromagnetic behavior. (**b**), MOKE signal from a bilayer CrI$_3$ showing vanishing Kerr rotation for applied fields between ±0.65 T, suggesting antiferromagnetic behavior. Insets depict bilayer magnetic ground states for different applied fields. (**c**), MOKE signal on a trilayer flake showing a return to ferromagnetic behavior.

# Supplementary Materials for

# Layer-dependent Ferromagnetism in a van der Waals Crystal down to the Monolayer Limit

Bevin Huang[1†], Genevieve Clark[2†], Efren Navarro-Moratalla[3†], Dahlia R. Klein[3], Ran Cheng[4], Kyle L. Seyler[1], Ding Zhong[1], Emma Schmidgall[1], Michael A. McGuire[5], David H. Cobden[1], Wang Yao[6], Di Xiao[4], Pablo Jarillo-Herrero[3*], Xiaodong Xu[1,2]*

[1]Department of Physics, University of Washington, Seattle, Washington 98195, USA
[2]Department of Materials Science and Engineering, University of Washington, Seattle, Washington 98195, USA
[3]Department of Physics, Massachusetts Institute of Technology, Cambridge, Massachusetts 02139, USA
[4]Department of Physics, Carnegie Mellon University, Pittsburgh, Pennsylvania 15213, USA
[5]Materials Science and Technology Division, Oak Ridge National Laboratory, Oak Ridge, Tennessee, 37831, USA
[6]Department of Physics and Center of Theoretical and Computational Physics, University of Hong Kong, Hong Kong, China

[†]These authors contributed equally to this work.
[*]Correspondence to: xuxd@uw.edu, pjarillo@mit.edu

**Content:**

**Extended Data Figures 1-4**

**S1: SQUID magnetometry on bulk samples**

**S2. Quantitaive optical microscopy in CrI$_3$**

**S3. Thin-film interferometry and the MOKE signal in CrI$_3$**

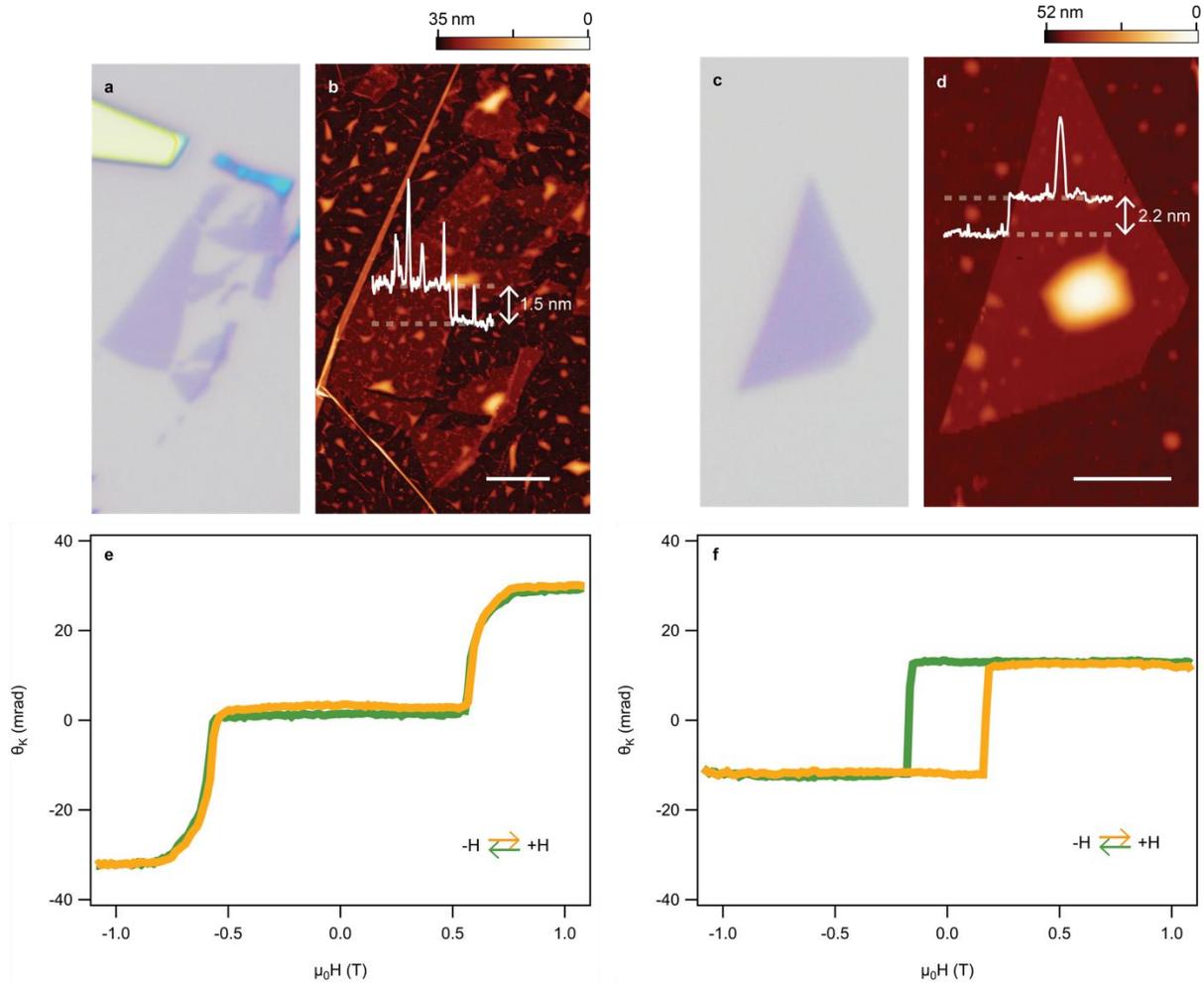

**Extended Data Figure 1 | Atomic force microscopy (AFM) and magneto-optic Kerr effect (MOKE) measurements of graphite-encapsulated few-layer CrI$_3$. a,** Optical microscope image of a bilayer CrI$_3$ flake on 285 nm SiO$_2$. **b,** AFM data for the CrI$_3$ flake in **a** encapsulated in graphite, showing a line cut across the flake with a step height of 1.5 nm. **c,** Optical microscope image of a tri-layer CrI$_3$ flake on 285 nm SiO$_2$. **d,** AFM data for the CrI$_3$ flake shown in **c** encapsulated in graphite. A line cut taken across the flake shows a step height of 2.2 nm. All scale bars are 2 μm in length. **e** and **f** show the MOKE signal as a function of applied magnetic field for the encapsulated bilayer in **b** and the encapsulated trilayer in **d** respectively.

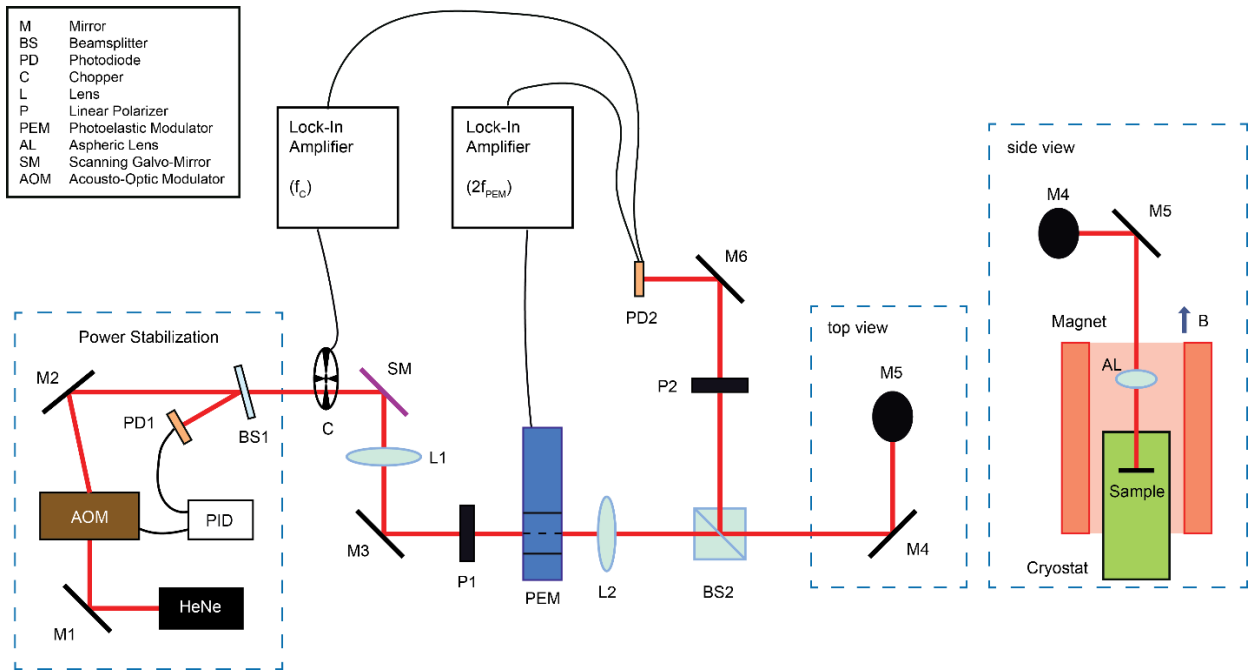

**Extended Data Figure 2 | Magneto-optical Kerr effect experimental setup.** Schematic of the optical setup used to measure Magneto-optical Kerr effect in $CrI_3$ samples. 633 nm optical excitation is provided by a power-stabilized HeNe laser. A mechanical chopper and photoelastic modulator provide intensity and polarization modulation, respectively. The modulated beam is directed through a polarizing beam splitter to the sample, housed in a closed-cycle cryostat at 15 K. A magnetic field is applied at the sample using a 7 T solenoidal superconducting magnet in Faraday geometry. The reflected beam passes through an analyzer onto a photodiode, where lock-in detection measures the reflected intensity (at $f_c$) as well as the Kerr rotation (at $f_{PEM}$).

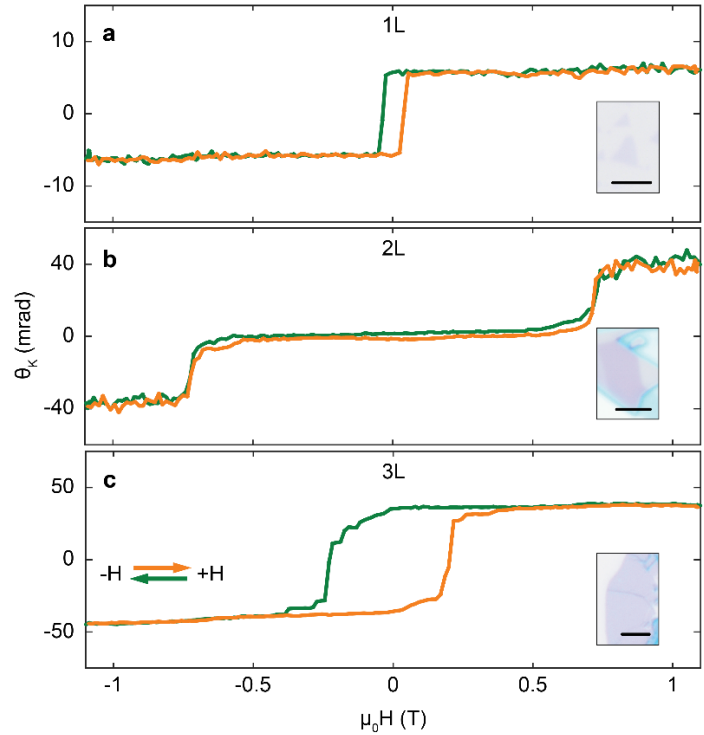

**Extended Data Figure 3 | Additional data for 1-3 layer samples showing Kerr rotation as a function of the applied magnetic field**. The insets show optical microscope images of the CrI$_3$ flakes at 100 x magnification. **a,** Additional monolayer data showing ferromagnetic hysteresis and remanent Kerr signal as a function of applied magnetic field. **b,** Additional bilayer data showing field-dependent behavior consistent with an antiferromagnetic ground state. **c,** Additional trilayer data showing field-dependent behavior consistent with a ferromagnetic ground state, as well as a larger remanent Kerr signal than the monolayer in **a**. All scale bars are 5 µm in length.

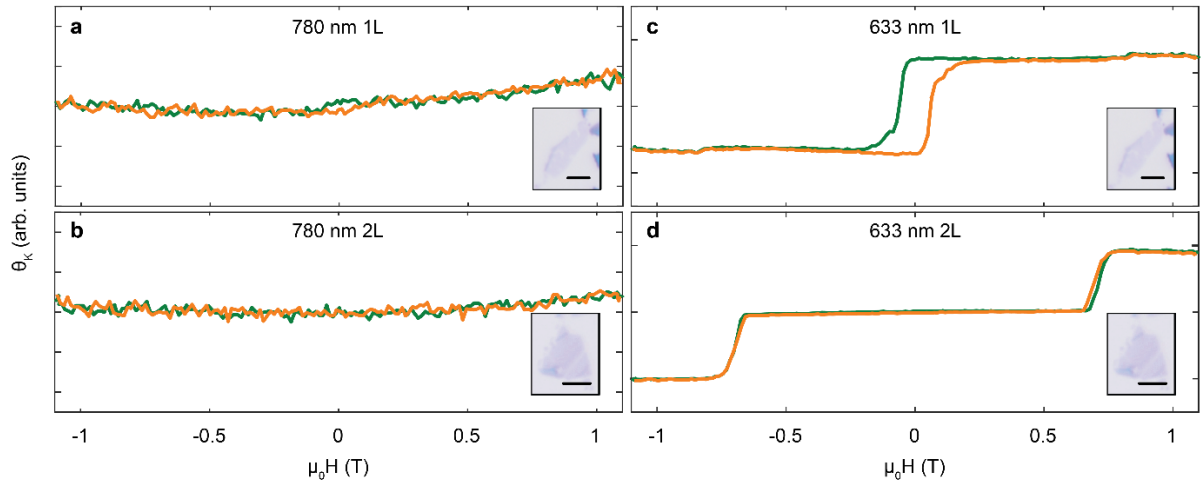

**Extended Data Figure 4 | Additional data for monolayer and bilayer samples under 780nm excitation and 633nm excitation.** Insets show optical microscope images of the CrI$_3$ flakes at 100x magnification. **a,** Kerr rotation as a function of applied magnetic field for a monolayer **a (c)**, and a bilayer **b (d)** under 780 nm (633 nm) excitation. All scale bars are 4 μm in length.

# Supplementary Information

## S1. SQUID magnetometry on bulk samples

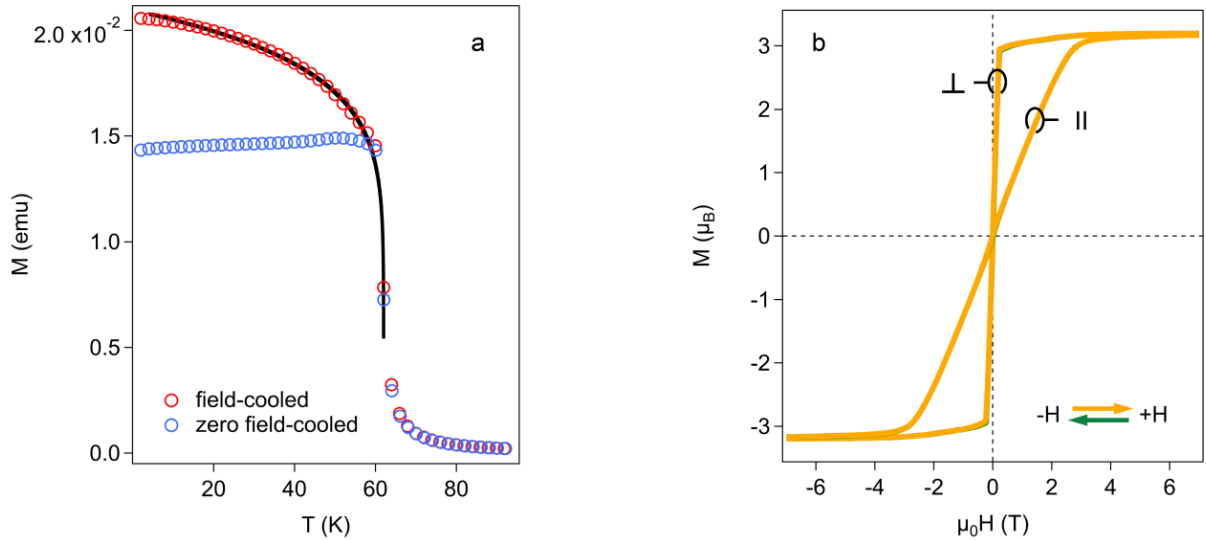

**Figure S1 | SQUID magnetometry in bulk CrI$_3$. a**, Zero-field cooled/ field-cooled temperature dependence of the magnetization of a CrI$_3$ bulk crystal with an applied magnetic field of 10 G perpendicular to the basal plane of the sample. The green line is a criticality fit ($M \propto (1-T_C/T)^\beta$) of the data with $T_C = 61$ K and $\beta = 0.125$ (Ising universality class). **b**, Hysteresis loops of the same sample with the external magnetic field in perpendicular ($\perp$) and parallel ($\parallel$) orientation with respect to the CrI$_3$ layers.

## S2. Quantitaive optical microscopy in CrI$_3$

Optical microscopy images were taken using a Nikon Eclipse LV-CH 150NA optical microscope with a DS-Ri2 full-frame camera. The setup was located inside a glove box (argon atmosphere) in order to prevent sample degradation. The quantitative optical contrast analysis required that images were captured with a 100x objective under monochromatic illumination at normal incidence. In practice, 10 nm full-width at half maximum (FWHM) filters were used (Andover Corp.) to filter the light coming from a halogen lamp.

The thickness of the flakes was determined by contact-mode atomic force microscopy (AFM) in ambient conditions. Given the extreme sensitivity of the samples to atmospheric moisture, the CrI$_3$ flakes were encapsulated between two pieces of few-layer graphite (typically 5 nm in thickness) prior to being extracted from the glove box.

For each optical microscopy filtered image, individual RGB values were extracted and averaged over each flake and substrate region to give the reflected intensities of the flake and substrate. The intensity was chosen to be exclusively the value of the channel with the highest number of counts. The experimental optical contrast value was then calculated according to the following expression (1).

$$C(d,\lambda) = \frac{I_{flake} - I_{substrate}}{I_{flake} + I_{substrate}} \quad (1)$$

Equation (1) expresses the relationship between the optical contrast $C$ between each flake and the substrate using the reflected intensities from the flake ($I_{flake}$) and the substrate ($I_{substrate}$). Figure 1 shows an example of a contrast map of a multi-step CrI$_3$ flake extracted from its optical micrograph.

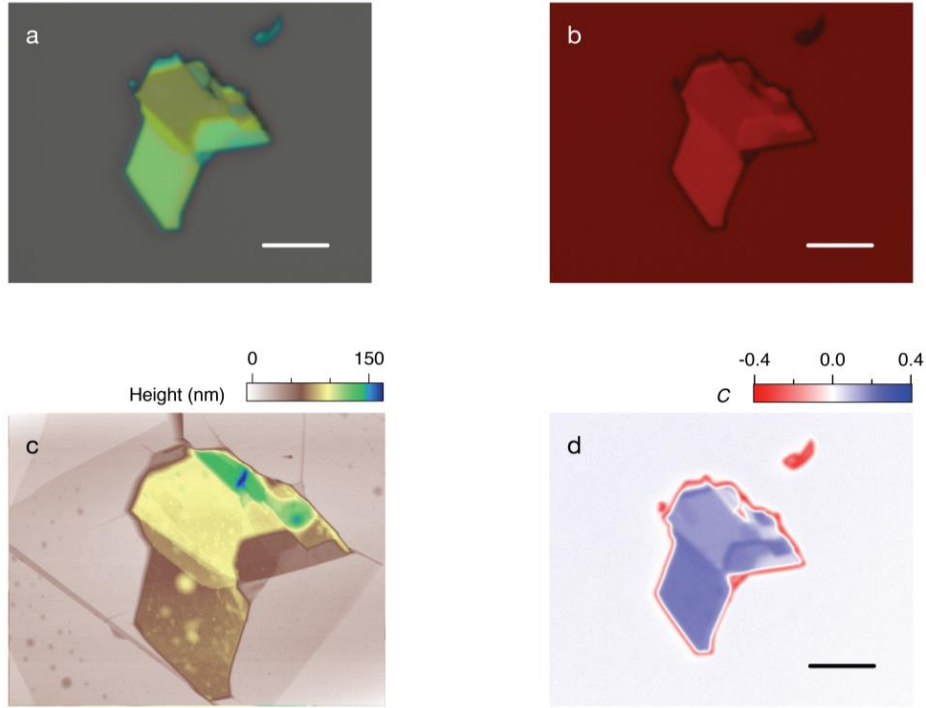

**Figure S2 | Thickness dependence of the optical contrast of the CrI$_3$ flakes.** Optical micrographs of a CrI$_3$ flake illuminated with white light (**a**). and with 631 nm (10 nm FWHM bandpass) filtered light (**b**). **c**, AFM topography image of the same sample. **d**, Optical contrast map extracted from the 631 nm micrograph in panel **b**. Scale bars are 5 µm long.

For flakes that have been exfoliated on a SiO$_2$/Si substrate, $C$ depends on the thickness of the flake and on the illumination wavelength [1,2]. Following the quantitative microscopy analysis proposed for graphene on SiO$_2$/Si substrates [3], $C$ can be computed for any kind of flake by using a model based on Fresnel's equations shown in (2a) and (2b):

$$I_{substrate}(\lambda) = \left| \frac{r_{02} + r_{23}e^{-2i\Phi_2}}{1 + r_{02}r_{23}e^{-2i\Phi_2}} \right|^2 \quad (2a)$$

$$I_{flake}(\lambda) = \left| \frac{r_{02}e^{i(\Phi_1+\Phi_2)} + r_{12}e^{-i(\Phi_1-\Phi_2)} + r_{23}e^{-i(\Phi_1+\Phi_2)} + r_{01}r_{12}r_{23}e^{i(\Phi_1-\Phi_2)}}{e^{i(\Phi_1+\Phi_2)} + r_{01}r_{12}e^{-i(\Phi_1-\Phi_2)} + r_{01}r_{23}e^{-i(\Phi_1+\Phi_2)} + r_{12}r_{23}e^{i(\Phi_1-\Phi_2)}} \right|^2 \quad (2b)$$

In this calculation, the subscripts 0, 1, 2, and 3 refer to air (treated as vacuum), $CrI_3$, $SiO_2$, and Si, respectively. The amplitude of the reflected path at the interface between media $j$ and $k$ is given by $r_{jk}$ in equation (3b) and is calculated from the complex refractive indices defined in equation (3a). $\Phi_j$ is the phase shift introduced by the interaction between light of wavelength $\lambda$ and medium $j$ with thickness $d_j$ shown in equation (3c).

$$\tilde{n}_j(\lambda) = n_j - i\kappa_j \quad (3a)$$

$$r_{jk} = \frac{\tilde{n}_j - \tilde{n}_k}{\tilde{n}_j + \tilde{n}_k} \quad (3b)$$

$$\Phi_j = \frac{2\pi \tilde{n}_j d_j}{\lambda} \quad (3c)$$

As can be noted from the previous expressions, if one wants to model $C$ using the Fresnel equations, the complex index of refraction of the material under study must be known. The reflectivity of $CrI_3$ measured in vacuum at 300 K at normal incidence of a single crystal platelet of $CrI_3$ has been previously reported [4]. This data was used to calculate the phase of the amplitude reflection coefficient $\theta$ at energies in the visible range by numerical integration, according to the Kramers-Kronig relation (4a). The refractive index $n$ and extinction coefficient $\kappa$ of $CrI_3$ were then obtained throughout the visible range by combining equations (4b) and (4c) at each energy value. The results are plotted in Figure 3.

$$\theta(E) = -\frac{E}{\pi} \int_0^\infty \frac{\ln[R(E')]}{(E')^2 - E^2} dE' \quad (4a)$$

$$r(E) = \sqrt{R(E)} e^{i\theta(E)} \quad (4b)$$

$$r(E) = \frac{n(E) - 1 + i\kappa(E)}{n(E) + 1 + i\kappa(E)} \quad (4c)$$

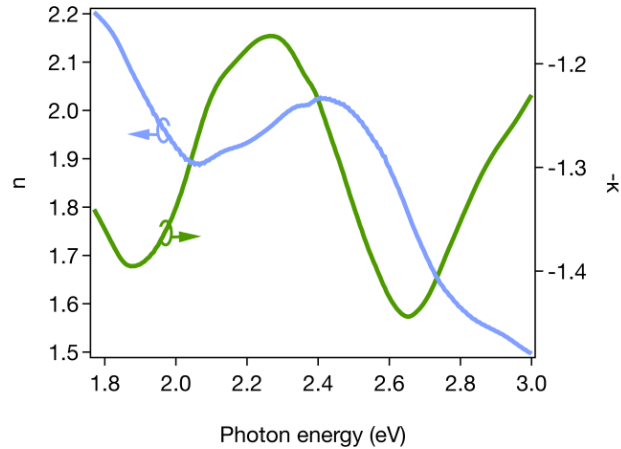

**Figure S3 | Computed index of refraction of bulk CrI$_3$.** Real ($n$) and imaginary ($k$) components are plotted as a function of photon energy in the visible range.

Substituting in for the complex indices of refraction of CrI$_3$, Si and SiO$_2$ [5] in equations (2a) and (2b), one can calculate the expected value of $C$ for flakes of different thicknesses as a function of the illumination wavelength. Figure 4(a) shows the contrast map for CrI$_3$ considering a fixed thickness of 285 nm of the SiO$_2$ layer in the SiO$_2$/Si substrate. We also present a line cut of that plot at an illumination wavelength of 635 nm in Fig. 4(b). It can be seen that the experimental data points follow closely the trend predicted by the model. Given that the method is non-destructive and can be performed inside a glove box for many different illumination wavelengths, the error in the determination of the number of layers can be reduced. This provides a fast and reliable method for the characterization of few-layer CrI$_3$ flakes.

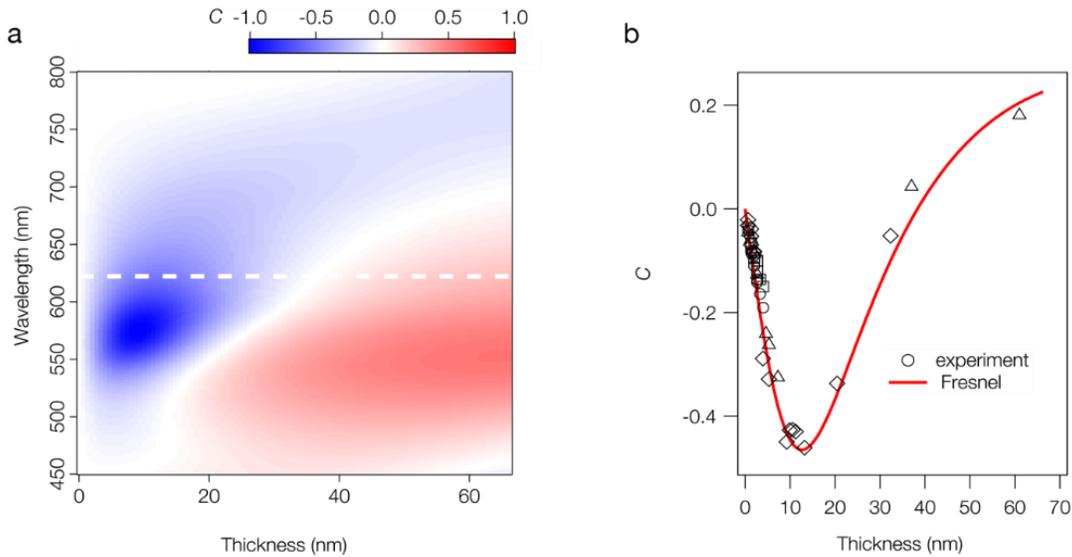

**Figure S4 | Fresnel model for the optical contrast ($C$) of CrI$_3$ flakes on Si/285 nm SiO$_2$ substrates. a**, Dependence of $C$ with the number of layers for a CrI$_3$ flake as a function of the illumination wavelength. **b**, Comparison of the experimental data with the computed thickness dependence of $C$ for a red-light-illuminated sample (line cut at 631 nm as shown by the dashed line in panel **a**). The different shape markers indicate data coming from different exfoliated samples.

## S3. Thin-film interferometry and the MOKE signal in CrI$_3$

Linearly polarized light is an equal superposition of right-circularly and left-circularly polarized light (RCP, LCP respectively). When a phase difference accrues between the RCP and LCP components, the polarization axis of the linearly polarized light rotates. This rotation can be observed in any material that exhibits circular birefringence. In magnetic samples, such as CrI$_3$, this birefringence arises from a non-zero magnetization, **M**, and is known as the magneto-optical Kerr effect (MOKE) when detected in reflection geometry. A MOKE measurement then detects changes in **M** by exploiting the functional dependence of the Kerr rotation on the magnetization,

$\theta_K(\mathbf{M})$. Additional interference terms must be accounted for however, when we discuss the Kerr rotation of a thin-film material, as reflections from the material-substrate ($CrI_3$-$SiO_2$) interface will superimpose with the reflection off the magnetic sample (fig. S5). As such, this motivates a model that uses the Fresnel equations (the same formalism from supplementary section 2) to calculate the reflection coefficients for RCP ($\tilde{r}_+$) and for LCP ($\tilde{r}_-$) light. The phase difference between $\tilde{r}_+$ and $\tilde{r}_-$ then, is the Kerr rotation:

$$\theta_K = \arg(\tilde{r}_+) - \arg(\tilde{r}_-). \quad (5)$$

To obtain the index of refraction for $CrI_3$ when fully spin polarized, we work in Cartesian coordinates and use a dielectric tensor of the form [6]:

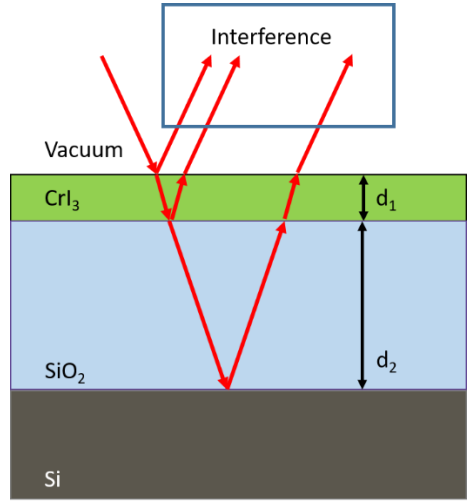

**Figure S5 | Thin-film interference ray diagram of $CrI_3$ on a silicon oxide/silicon substrate.** Light incident on $CrI_3$ undergoes reflections at the $CrI_3$-$SiO_2$ interface (green-blue boundary) as well as the $SiO_2$-Si interface (blue-gray boundary). These reflections interfere with the initial reflection off the $CrI_3$ flake to produce thin-film interference that depends on the $CrI_3$ layer thickness, $d_1$, as well as the $SiO_2$ thickness, $d_2$. The underlying silicon wafer is assumed to be semi-infinite.

$$\varepsilon = \begin{bmatrix} \tilde{\varepsilon}_{xx} & iQM & 0 \\ -iQM & \tilde{\varepsilon}_{xx} & 0 \\ 0 & 0 & \tilde{\varepsilon}_{zz} \end{bmatrix}, \quad (6)$$

where Q is the complex Voigt constant and M is the out-of-plane (parallel to the z-axis) component of the magnetization, which we assume is constant for all $CrI_3$ thicknesses when fully spin polarized. This form of the dielectric tensor for a magnetic sample is valid assuming: *i.* the crystal exhibits at least three-fold symmetry; *ii.* **M**, is parallel to the axis of rotation that gives rise to the three-fold symmetry; and *iii.* the axis of rotation is chosen to be in the z-axis. Solving for the normal modes, we get the eigenvalues:

$$\tilde{n}_\pm = \sqrt{\tilde{\varepsilon}_{xx} \pm QM} \quad (7)$$

and the eigenvectors to be:

$$\mathbf{D}_\pm = (E_x \pm iE_y)\hat{\mathbf{z}} \quad (8)$$

where, $\tilde{n}_+$, $\mathbf{D}_+$ denote the eigenvalue and eigenvector respectively of RCP light and $\tilde{n}_-$, $\mathbf{D}_-$ denote the eigenvalue and eigenvector respectively of LCP light in CrI$_3$. The complex dielectric component, $\tilde{\varepsilon}_{xx}$, when related to the $\tilde{n}$ defined in (3a), is:

$$\tilde{\varepsilon}_{xx} = n^2 - \kappa^2 - 2in\kappa, \quad (9)$$

where the value of $\tilde{\varepsilon}_{xx}$ was derived from the $n$ and $\kappa$ at ~1.96 eV modeled in figure S3. It is apparent from (7) that we should not expect $\theta_K$ to depend linearly on **M** and layer number. In addition, interference from reflections off the CrI$_3$-SiO$_2$ interface will give rise to a non-trivial functional form of $\theta_K$ with respect to layer thickness.

There is no determination of Q in the literature for CrI$_3$, so we varied Q as a complex parameter and constrained it to a small range that fit experimental $\theta_K$ values from our MOKE measurements on trilayer and bulk CrI$_3$. These calculations qualitatively describe the large increase in $\theta_K$ from monolayer to trilayer, as well as the negative $\theta_K$ seen at positive $\mu_0$H for bulk flakes. However this simple model does not incorporate layer-dependent electronic structure changes, seen in other atomically-thin van der Waals materials such as MoS$_2$ [7], and evident in this system as a change in magnetic ground states from monolayer to bilayer.